\documentclass[conference]{IEEEtran}
\IEEEoverridecommandlockouts
\usepackage{amsmath,amssymb,amsfonts}
\usepackage{algorithmic}
\usepackage{graphicx}
\graphicspath{ {./figures/} }
\usepackage{textcomp}
\usepackage{xcolor}
\usepackage{flushend}
\def\BibTeX{{\rm B\kern-.05em{\sc i\kern-.025em b}\kern-.08em
		T\kern-.1667em\lower.7ex\hbox{E}\kern-.125emX}}

\usepackage[citestyle=numeric,style=numeric,backend=bibtex,sorting=none]{biblatex}
\newcommand{\myvec}[1]{\ensuremath{\vec{#1}}}

\begin{document}

	\title{Towards Automated Metamorphic Test Identification for Ocean System Models}
	
	\author{\IEEEauthorblockN{1\textsuperscript{st} J Hiremath, Dilip }
		\IEEEauthorblockA{\textit{MarDATA -- Helmholtz School for Marine Data Science} \\
			\textit{GEOMAR Helmholtz Centre for Ocean Research Kiel} \\
			\textit{\& Kiel University}\\
			Kiel, Germany \\
			0000-0001-7095-6583}
		\and
		\IEEEauthorblockN{2\textsuperscript{nd} Claus, Martin}
		\IEEEauthorblockA{\textit{Ocean Circulation and Climate Dynamics Department} \\
			\textit{GEOMAR Helmholtz Centre for Ocean Research Kiel}\\
			\textit{\& Kiel University}\\
			Kiel, Germany \\
			0000-0002-7525-5134}
		\and
		\IEEEauthorblockN{3\textsuperscript{rd} Hasselbring, Wilhelm}
		\IEEEauthorblockA{\textit{Software Engineering Group, Computer Science Department} \\
			\textit{Kiel University}\\
			Kiel, Germany \\
			0000-0001-6625-4335}
		\and
		\IEEEauthorblockN{4\textsuperscript{th} Rath, Willi}
		\IEEEauthorblockA{\textit{Ocean Circulation and Climate Dynamics Department} \\
			\textit{GEOMAR Helmholtz Centre for Ocean Research Kiel}\\
			Kiel, Germany \\
			0000-0003-1951-8494}
	}
	
	\maketitle
	
	\begin{abstract}
		
		Metamorphic testing seeks to verify software in the absence of test oracles.
		Our application domain is ocean system modeling, where test oracles rarely exist, but where symmetries of the simulated physical systems are known.
		The input data set is large owing to the requirements of the application domain.

		This paper presents work in progress for the automated generation of metamorphic test scenarios using machine learning.
		We extended our previously proposed method~\cite{Hiremath2020} to identify metamorphic relations with reduced computational complexity.
		Initially, we represent metamorphic relations as identity maps. 
		We construct a cost function that minimizes for identifying a metamorphic relation orthogonal to previously found metamorphic relations and penalize for the identity map.
		A machine learning algorithm is used to identify all possible metamorphic relations minimizing the defined cost function.
		We propose applying dimensionality reduction techniques to identify attributes in the input which have high variance among the identified metamorphic relations. 
		We apply mutation on these selected attributes to identify distinct metamorphic relations with reduced computational complexity. 
		For experimental evaluation, we subject the two implementations of an ocean-modeling application to the proposed method to present the use of metamorphic relations to test the two implementations of this application.
		
	\end{abstract}
	
	
	\section{Introduction}
	
	Software quality assurance forms an important part of the software development cycle.
	Its importance is accepted throughout the software engineering community.
	Verification and validation of an application are essential steps in software testing.
	The increasing complexity of present-day software and the use of machine learning techniques presents new challenges to test software sufficiently.
	However, machine learning techniques are used in research and in developing solutions for real-world applications \cite{Reichstein2019,EcoInf2017}.  
	Besides this, the nature of research software is exploratory.
	The output for such software is usually unknown and cost-intensive to compute.
	The unknown output of the application is known as the test oracle problem in software engineering.
	
	There have been many approaches to test software that pose the test oracle problem.
	Metamorphic testing (MT) is the most prominent approach among these, as seen by the increase in number of papers published and increased research efforts in recent times.
	MT was first introduced by \cite{Chen1998}.
	It has seen an increase adoption in research and real-world applications.
	Comprehensive surveys of metamorphic testing may be found in~\cite{Zhang2020,Segura2020,Lin2020,Segura2019a,Segura2018,Segura2016}.
	
	Metamorphic testing is based on the idea that most of the time it is easier to predict relations between outputs of a program, than understanding its input-output behavior \cite{Saha2020}.
	The central element of metamorphic testing is the metamorphic relation (MR). An MR is a necessary property of the target application in relation to multiple inputs and their corresponding outputs \cite{Chen2018b}.
	
	Identifying metamorphic relations is labor-, time-, and cost-intensive.
	Experienced domain experts require a thorough understanding of the application and the desired behavior of MRs to identify metamorphic test scenarios.
	Automating this process would reduce the cost and increase the probability of widespread adoption.
	Using the available automated test case generation methods, we can develop test cases once the MRs are available as shown by Hui et al.~\cite{Hui2020}.
	We can harness testing frameworks to automate application testing using the generated test cases and test suites.
	
	MT is not without its limitations.
	A major limit is that MT by itself cannot prove that the output of the application is correct~\cite{Segura2019a}.
	It is also challenging to quantify the effectiveness of MT.    
	Since the basis of MT relies on comparing multiple successive outputs with respect to each other and the morphed inputs. If the initial outputs themselves were wrong and the successive outputs were within an expected variation satisfying the MR, MT would not detect the error in the software under test.
	The other limitation of the MT approach is that for comprehensive testing the number of MRs required is large and thus leads to an extensive test suite.
	MT finds faults in the application but does not point to the function with the bug, thus debugging the software under test is challenging.
	One method is to follow the changes to input parameters for failed metamorphic tests, find the parameters with high covariance, and investigate their path for debugging.
	
	Section~\ref{motivation} explains the motivation behind our research. 
	In Section~\ref{s-application} we introducing an example application, before formulating the problem to be solved in Section~\ref{s-problem}. 
	In Section~\ref{s-method}, we explain our proposed method and sketch the solution employing machine learning. 
	Section~\ref{s-future}  of the paper highlights the challenges and an outlook to future work. 
	Section~\ref{data_availability}  provides the details of the resources available for this paper. 
	
	\section{Motivation}\label{motivation}
	
	Our application domain are ocean system models.
	Currently, there are no papers based on MT focused on ocean modeling, to the best of our knowledge.
	However, there are a few papers on applying MT to geographic information systems \cite{Lin2020,Hui2020}
	Ocean system modeling software is often based on legacy code that has over the years undergone multiple development cycles leading to complex intertwined code.
	They rarely employ a standard systematic testing approach to verify the software.
	The challenges to this have been explained by \cite{Johanson2018}, \cite{Kanewala2014b}, and \cite{Kelly2008}.
	The focus is on assimilating functions that work rather than developing the software adhering to the principles of software engineering.
	Verification is based on the modeler's undocumented plausibility check to see if the output lies within the output range, the researcher is expecting.
	Though researchers accept software testing as a value-adding step in developing software, they rarely identify or maintain separate test cases to verify their functions and software.
	
	Some recent developments of ocean simulation software are collaborative, open-source, and in newer programming languages.
	This has led to refactoring some existing components and modular addition of new functions to the existing codebase.
	Testing efforts are focused on regression tests as part of the continuous integration framework in most collaborative open-source scientific software development~\cite{OSRS2020}. A goal should be to follow the FAIR principles not only for research data but also for research software~\cite{FAIR_Software_2020}.

	The increasing use of machine learning techniques to compute intermediate input data points demands non-traditional software techniques to ensure the correctness of the software in use.
	There is a substantial gap between adopting artificial intelligence in software development and testing capabilities for valid implementations of artificial intelligence methodologies.
	Testing these systems has proven to be challenging.
	Though there is an increase in the solution's effectiveness by adopting these techniques, the lack of methods to verify and the inability to explain the improvement in effectiveness has resulted in cautious adoption of these software development techniques.
	With the increasing assimilation of artificial Intelligence methodologies as a part of software engineering, the need for better testing approaches for validating software posing test oracle problems has increased.
	
	From our current knowledge, there does not exist an established standard testing procedure for verifying the developed scientific software.
	Most of the verification efforts are based on unit tests written, managed, and executed by contributors at their discretion.
	A structured quality assurance framework would help with the adoption of the testing capabilities across the research facility and lead to more reliable software development.
	One of the major hurdles to adopting MT is the cost associated with identifying metamorphic relations.
	The central idea of this short paper is to present an approach to automate identifying metamorphic tests.
	This will increase the widespread application of MT as an efficient verification technique for the software that poses test oracle problems.
	
	The future of Software Engineering will revolve around harnessing artificial intelligence techniques for solving real-world complex and interdependent problems.
	Automated identification of metamorphic tests will lead to improved adoption of MT-based test suit development for automated program repair as shown by Jiang et al.~\cite{Jiang2020}.

	\section{Example Application}\label{s-application}
	
	We use a simple but realistic ocean-modeling application to show our approach~\cite{Rath2019}.
	We can access the application on Binder ~\cite{Jupyter2018BinderScale.}.
	The example application is to calculate a time series of Kinetic energy of the surface ocean for randomly generated data for a grid of ($10 \times 20$) for $30$ temporal resolutions
	\begin{equation}
		\label{eq:energy}
		e(t, y, x) = \frac{1}{2}\frac{\int{\rm d}y{\rm d}x (u(t, y, x)^2 + v(t, y, x)^2)}{\int{\rm d}y{\rm d}x}
	\end{equation}
	where $e$ represents the time series of Kinetic energy, $t$ the time, $y$ and $x$ are spatial coordinates, and horizontal surface-velocities $u$ and $v$ are calculated from the sea-level $\eta$ using
	\begin{equation}
		\label{eq:uv}
		(u, v)(t, y, x) = \frac{G}{F} \left(-\frac{\partial}{\partial y}, \frac{\partial}{\partial x}\right) \eta(t, y, x)
	\end{equation}
	where $G$ represents the gravitational acceleration (in $meters/second^2$) and $F$ the Coriolis parameter (in $1/second$) at $30$\textdegree\ North.
	
	The two functions of Equation~\eqref{eq:energy} are coded in the sample application, one of which is not respecting cyclic boundary conditions.
	To numerically implement Equation~\eqref{eq:energy}, we discretize the integral in \eqref{eq:energy} and the partial derivatives in \eqref{eq:uv} by applying the method of finite differences and link it back to Equation~\eqref{eq:metamorphic_relation}.
	Thus, the input data $\myvec{x}$ contains all discretized values of sea level $\eta$, of all coordinates $t$, $y$, $x$, and the physical constants $G$ and $F$.
	A metamorphic transformation $g(x)$ could change all or some of the atomic data points in $\myvec{x}$ such that all $e(g(\myvec{x})) = e(\myvec{x})$ for all discrete values of time~$t$.
	
	\begin{figure*}[htb]
		\centering
		\includegraphics[width=0.7\textwidth]{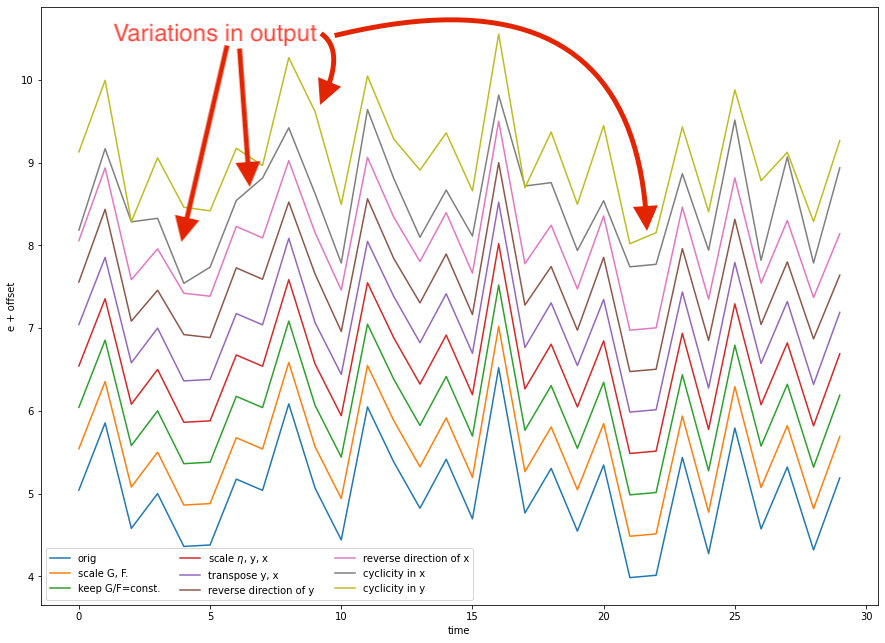}
		\caption{Time series outputs of Kinetic energy of surface ocean calculated using the function not respecting the cyclic boundary conditions for (x,y) coordinates for multiple applied MRs.}
		\label{fig:appied-MRs}
	\end{figure*}

	Fig.\ref{fig:appied-MRs} shows the time series output for multiple applied MRs calculated using the function in which the cyclic boundary condition is not implemented.
	These MRs were manually identified based on the symmetries of the energy equation. 
	The bottom most plot is that of the original input data.
	The applied MRs are multiplying $g$ and $h$ by the same constant, Keeping $G/F$ as constant, scaling $\eta, x$ and $y$ by a constant, transposing $x$ and $y$, reversing the direction of $x$, reversing the direction $y$, transposing the $x$ coordinates cyclically and transposing the $y$ coordinates cyclically.
	We observe that for applied initial MRs the time series outputs are in accordance with the applied MRs.
	Time series out puts for transposing the $x$ coordinates cyclically and transposing the $y$ coordinates cyclically differs from the rest as highlighted in Fig.~\ref{fig:appied-MRs} by red arrows.

	\section{Problem Formulation}\label{s-problem}
	
	For the given application domain, we may define the function under test $f: X \to Y$, where both $X$ and $Y$ are finite dimensional normed vector spaces.
	Then a metamorphic relation $(g, h)$ satisfies
	\begin{equation}
		\label{eq:metamorphic_relation}
		f(g(\myvec{x})) = h(f(\myvec{x}))
	\end{equation}
	where $\myvec{x} \in X$, $g: X \to X$ is the input mapping from source input to follow-up (morphed) input, and $h: Y \to Y$ is the output mapping from the source output to the follow-up output, that is the output associated with the morphed input.
	
	Throughout this paper, we limit $h$ to be the identity map
	\begin{equation}
		\label{eq:metamorphic_relation_identiy_assumption}
		f(g(\myvec{x})) = f(\myvec{x})
	\end{equation}
	and we constrain $g$ to be an affine transformation
	\begin{eqnarray}
		\label{eq:affine_transformation}
		g(\myvec{x}) = \Gamma\cdot\myvec{x} + \myvec{\beta}
	\end{eqnarray}
	where $\Gamma$ is a matrix associated with an endormophism of $X$, and $\myvec{\beta} \in X$ is an offset.
	Note that the limitation of $g$ being an affine transformation is motivated by retaining the ability of capturing physical properties related to symmetries of the governing equations, such as invariance under translation or rotation, while greatly reducing the soultion space.

	\section{Proposed Method}\label{s-method}

	There may exist an infinite number of possible $g$ satisfying \eqref{eq:metamorphic_relation_identiy_assumption} but one always is the identity map.
	The task is then to find new $g$, or at least approximations to $g$, which differ from all $g$ known so far.
	For this we use an iterative approach.
	We represent the set of identified metamorphic relations associated with $f$ as
		\begin{equation}
		\label{eq:g}
		G^f_{n-1} = \{g_0, g_1, \ldots, g_{n-1}\}
	\end{equation}

	We seek to find a new possible $g_n$ by minimizing a cost function that rewards a $g_n$ which minimizes the distance between source output and follow-up output $|f(g_n(\myvec{x})) - f(\myvec{x})|$ and penalizes if $g_n$ is already known.
	Throughout the text, we take the norm $\left|\cdot\right|$ to be the Euclidean distance.
	The cost function for a candidate $g_n$ and a given $G^f_{n-1}$ is defined by
	\begin{equation}
		\label{eq:cost_function}
		J(g_n, G^f_{n-1}) = \int_{X}\frac{|f(g_n(\myvec{x})) - f(\myvec{x})|}{\epsilon + \prod_{g \in G^f_{n-1}}\left|g_n(\myvec{x}) - g(\myvec{x})\right|^2} dx
	\end{equation}
	where $g_n$ represents a potential morphing to $\myvec{x}$ such that it satisfies~\eqref{eq:metamorphic_relation_identiy_assumption} and $\epsilon$ is a small machine precision constant to avoid division by zero when $g_n$ is identical to a previously identified $g \in G^f_{n-1}$. 
	Note that in the denomiator we write the product of squared differences to ensure $\lim_{g_n \to g} J(g_n, G^f_{n-1}) = \infty$ which heavily penalizes $g_n$ for being close to an element of $G^f_{n-1}$.

	We apply a modified algorithm loosely based on the genetic algorithm with a Monte Carlo optimization to find a $g_n$ that minimizes the cost function~\eqref{eq:cost_function}.
	We start the iteration with $G^f_0 = \{\mathrm{id}_X\}$ containing only the identity map on $X$.
	Initially, we assign random values to $\{\Gamma, \myvec{\beta}\}$ and in each step a mutation $\{\delta\Gamma, \myvec{\delta\beta}\}$ is proposed.
	An acceptance threshold $p$ is set which determines at what rate a mutation that would increase the cost function is still not rejected.
	If the cost after the application of the mutation is smaller than before the mutation, the applied mutation is always accepted.
	The process is repeated to move towards a global minimum.
	After converging to a minimum, we update the solution set $G^f_n = G^f_{n-1} \cap \{g_n\}$. 
	Then a new $g_{n+1}$ is created with randomly assigned $\{\Gamma, \myvec{\beta}\}$ and the process is restarted, this time minimizing $J(g_{n+1}, G^f_n)$.

	With this method we are able to identify multiple metamorphic transformations which satisfy equation \eqref{eq:metamorphic_relation_identiy_assumption}.

	\section{Challenges and Future Work}\label{s-future}
	
	We identify multiple MRs using the approach proposed in Section \ref{s-method}.
	We suspect the identified MRs are a chain of multiple elementary constituent MRs.
	Hence, it will be challenging to explain identified MRs in terms of physical symmetries that hold in the real oceans.
	We propose dimensionality reduction techniques to decompose the identified chained MRs.
	The decomposed MRs will be closer to the constraints of the Physical Ocean and easier to interpret.
	Since the resulting MRs will be distinct after decomposing, we expect test cases constructed with these MRs to be effective in uncovering defects.
	They will also increase the test coverage of the application.
	
	Since we assume input and output mapping relations, MRs corresponding to functional states of the software are not identified by the method proposed in Section \ref{s-method}.
	
	Further work is to relax the restriction to $h$ in~\eqref{eq:metamorphic_relation} being the identity map.
	Next steps will explore other optimization methods for the cost function constructed in Section\ref{s-method} including deep learning frameworks.

	\section{Data Availability}\label{data_availability}
	
	The sample application presented in Section~\ref{s-application} and the data generated are available in \cite{Rath2019}.
	The published artifact contains a link to an interactive session where the example application can be executed.

	\section*{Acknowledgments}\label{affiliations}
	
	The first author is funded through the Helmholtz School for Marine Data Science (MarDATA), Grant No. HIDSS-0005.
	We have developed the approach in collaboration of the Computer Science Department at Kiel University and the Ocean Circulation and Climate Dynamics Department at the GEOMAR Helmholtz Centre for Ocean Research Kiel.

	\printbibliography 
	
\end{document}